\documentclass[twocolumn]{emulateapj}
\usepackage{ amsmath, color, natbib}%amsmath, amsfonts,amssymb,natbib,graphicx,color,placeins}
\newcommand{\kms}{km~s$^{-1}$}

\begin{document}
\title{Characterizing spiral arm and interarm star formation}

\author{K. Kreckel\altaffilmark{1},  G. A. Blanc\altaffilmark{2,3,4}, E. Schinnerer\altaffilmark{1}, B. Groves\altaffilmark{5}, A. Adamo\altaffilmark{6}, A. Hughes\altaffilmark{7,8}, S. Meidt\altaffilmark{1}}

\altaffiltext{1}{Max Planck Institut f\"{u}r Astronomie, K\"{o}nigstuhl 17, 69117 Heidelberg, Germany;  kreckel@mpia.de}
\altaffiltext{2}{Departamento de Astronom'a, Universidad de Chile, Camino del Observatorio 1515, Las Condes, Santiago, Chile}
\altaffiltext{3}{Centro de Astrof'sica y Tecnolog'as Afines (CATA), Camino del Observatorio 1515, Las Condes, Santiago, Chile}
\altaffiltext{4}{Visiting Astronomer, Observatories of the Carnegie Institution for Science, 813 Santa Barbara St, Pasadena, CA, 91101, USA}
\altaffiltext{5}{Research School of Astronomy and Astrophysics, Australian National University, Canberra, ACT 2611, Australia}
\altaffiltext{6}{Department of Astronomy, The Oskar Klein Centre, Stockholm University, AlbaNova University Centre, SE-106 91 Stockholm, Sweden}
\altaffiltext{7}{CNRS, IRAP, 9 Av. du Colonel Roche, BP 44346, F-31028 Toulouse cedex 4, France}
\altaffiltext{8}{Universit\'{e} de Toulouse, UPS-OMP, IRAP, F-31028 Toulouse cedex 4, France}

\begin{abstract}
Interarm star formation contributes significantly to a galaxy's star formation budget, and provides an opportunity to study stellar birthplaces unperturbed by spiral arm dynamics.  
Using optical integral field spectroscopy of the nearby galaxy NGC 628 with VLT/MUSE, we construct H$\alpha$ maps including detailed corrections for dust extinction and stellar absorption to identify 391 HII regions at 35pc resolution over 12 kpc$^2$.  Using tracers sensitive to the underlying gravitational potential, we associate HII regions with either arm (271) or interarm (120) environments.    
Using our full spectral coverage of each region, we find that most HII region physical properties (luminosity, size, metallicity, ionization parameter) are independent of environment.
We calculate the fraction of H$\alpha$ luminosity due to the diffuse ionized gas (DIG) background contaminating each HII region, and find the DIG surface brightness to be higher within HII regions compared to the surroundings, and slightly higher within arm HII regions.  
Use of the temperature sensitive [SII]/H$\alpha$ line ratio map instead of the H$\alpha$ surface brightness to identify HII region boundaries does not change this result.  Using the dust attenuation as a tracer of the gas, we find depletion times consistent with previous work ($2 \times 10^9$ yr)  with no differences between the arm and interarm, however this is very sensitive to the DIG correction.  Unlike molecular clouds, which can be dynamically affected by the galactic environment, we see fairly consistent HII region properties in both arm and interarm environments. 
This suggests either a difference in arm star formation and feedback, or a decoupling of dense star forming clumps from the more extended surrounding molecular gas. 
\end{abstract}

%>>>>>>>>>>>>>>>>>>>>>>>>>>>>>>>>>Section<<<<<<<<<<<<<<<<<<<<<<<<<<<<<<<<<<<<<<<<<<<<
\section{Introduction}

Studies of extragalactic star formation are necessarily biased towards studying spiral arms, where both gas and star formation are concentrated.  However, giant molecular clouds spend as long or even longer in the interarm \citep{Dobbs2006}, and a significant fraction (30-60\%) of star formation in nearby galaxies is observed to occur in interarm regions \citep{Foyle2010}.  This fraction drops (8\% in M51, \citealt{Lee2011}; 14\% in M31, \citealt{Azimlu2011}) in studies that resolve individual HII regions, where the brightest HII region is almost an order of magnitude brighter in the arm compared to the interarm.  Some of this difference may be due to contamination of star formation tracers on large ($\sim$500 pc) scales by emission from diffuse ionized gas (DIG), which makes up on average $\sim$60\% of the total H$\alpha$ flux of a galaxy \citep{Rand1992, Greenawalt1998, Zurita2000, Oey2007, Haffner2009}.  

Most existing studies comparing spiral arm and interarm HII regions are performed using H$\alpha$ narrow-band imaging, and as such they are limited to an analysis of the HII region luminosities and sizes without the ability to properly correct for varying contribution from [NII], dust extinction or the DIG.  Statistical studies of HII regions have found evidence for a steeper slope in the luminosity function for the interarm compared to the spiral arms \citep{Kennicutt1980, Kennicutt1989HII, Rand1992, Thilker2000, Scoville2001}, however in many cases no difference has been observed \citep{Knapen1993, Rozas1996, Knapen1998, Gutierrez2011}.  The changing slope is largely attributed to statistical effects, as spiral arms contain more HII regions and therefore more luminous regions \citep{Knapen1998}, or increased blending of individual HII regions, as the arms have an overall higher surface density \citep{Scoville2001}.  In their detailed study of HII regions in M31, \cite{Azimlu2011} found similar slopes at the bright-end, but that the luminosity functions for arm and interarm regions peak at different luminosities, which they attribute to a large population of aged B stars in the interarm from a recent starburst.   

Recent development of new optical integral field unit (IFU) spectrographs allows for measurement of intrinsic H$\alpha$ emission from HII regions, including corrections for [NII], dust attenuation, stellar absorption, and DIG,  while providing full spectral and 2D information.  While most of these effects are directly constrained by well established methods using the additional spectral information, measuring the amount of H$\alpha$ flux arising from the DIG is more uncertain.  In addition to spatial identification via its more diffuse structure \citep{Zurita2000}, the DIG also exhibits distinctive spectral diagnostic line ratios \citep{Madsen2006, Blanc2009}. As the DIG morphology is expected to be complex, and vary across environmental features, deep spectral mapping is necessary to rigorously isolate this contaminant to star formation studies.
  
There are many large optical IFU galaxy surveys underway with kpc scale resolution (CALIFA, \citealt{Sanchez2012}; MANGA, \citealt{Bundy2015}; SAMI, \citealt{Bryant2015}), well suited to study bulk star formation in galaxies. The tight correlation between molecular gas and star formation on these scales suggests a  universal star formation law  \citep{Schmidt1959, Kennicutt1989, Kennicutt1998, Bigiel2008}, however detailed studies reveal an additional dependence on local conditions and environmental parameters \citep{Leroy2013}.  As most HII regions have sizes between 10-100~pc \citep{Azimlu2011, Gutierrez2011, Whitmore2011}, significant information is lost with spatial averaging.   Nearby galaxies provide the only opportunity to explore the physics of star formation at the relevant $<$50pc scales.  

We explore the impact of spiral arms on galaxy evolution by fully characterizing the physical conditions of arm and interarm HII regions using VLT/MUSE optical IFU spectroscopy within the nearby galaxy NGC 628 at 35pc scales.  While our imaging does not cover the full disk of the galaxy, it provides a first look at the statistical studies enabled by these new observing techniques, and allows us to explore the use of diagnostic line ratios to fully characterize HII region properties (metallicity, ionization parameter, DIG fraction).  NGC 628, a near face-on type SAc grand design spiral, has a global star formation rate (SFR) of 0.68 M$_\sun$ yr$^{-1}$ \citep{Calzetti2010}, and a stellar mass 3.6$\times10^9 M_\sun$ \citep{Skibba2011}.  We assume a  distance of 7.2 Mpc (1$\arcsec=35$pc) and R$_{25}$=5\farcm25 \citep{Kennicutt2011}.  After a brief description of the data (Section~2) and our derived HII region properties (Section~3), we quantify differences in these properties between arm/interarm including DIG contribution (Section~4.1 \& 4.2) and discuss the impact on the star formation law (Section~4.3) before summarizing in Section~5.

%>>>>>>>>>>>>>>>>>>>>>>>>>>>>>>>>>Section<<<<<<<<<<<<<<<<<<<<<<<<<<<<<<<<<<<<<<<<<<<<
\section{Data}
\begin{figure*}
\centering
\includegraphics[width=6.5in]{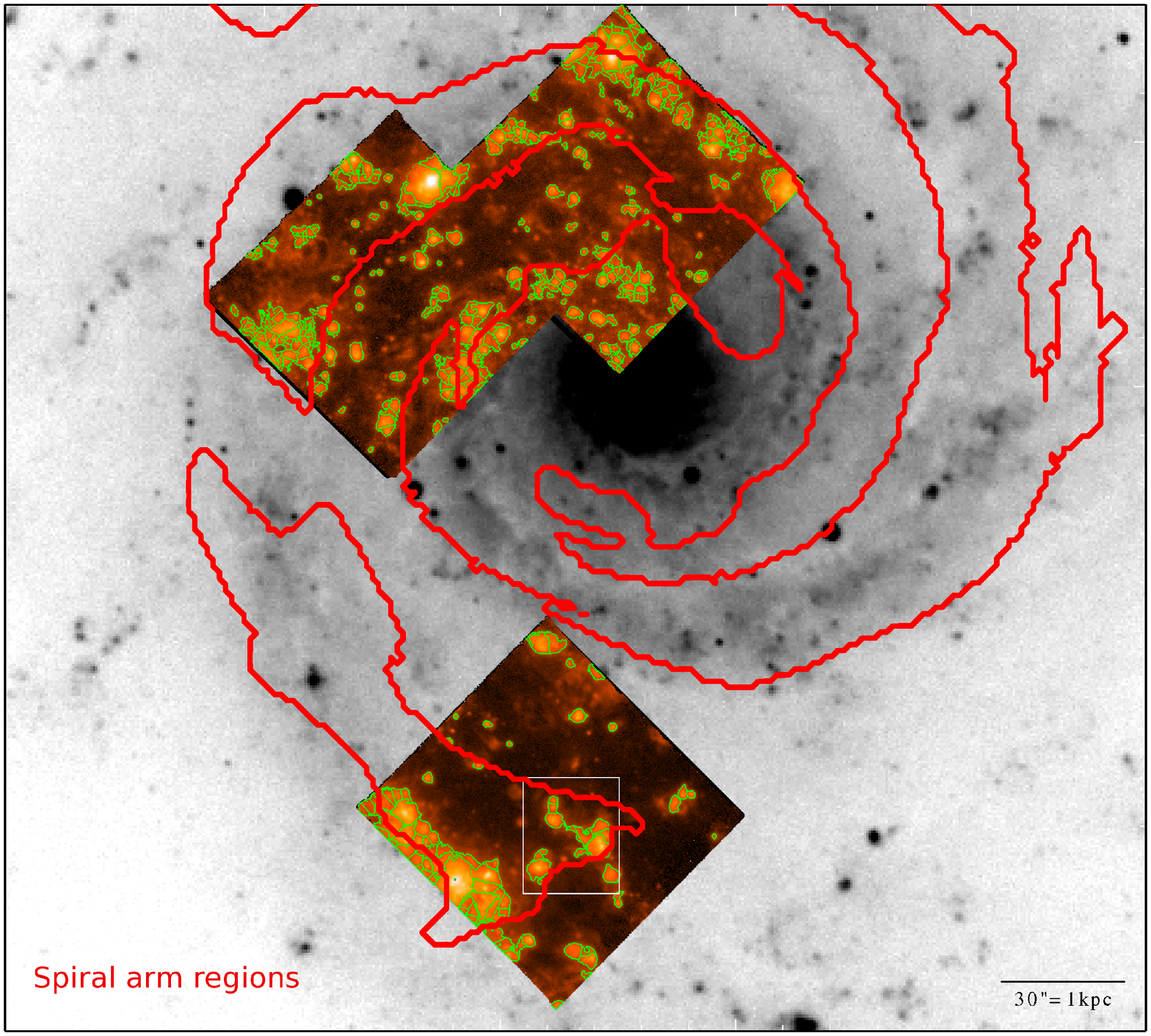}
\includegraphics[width=6.8in]{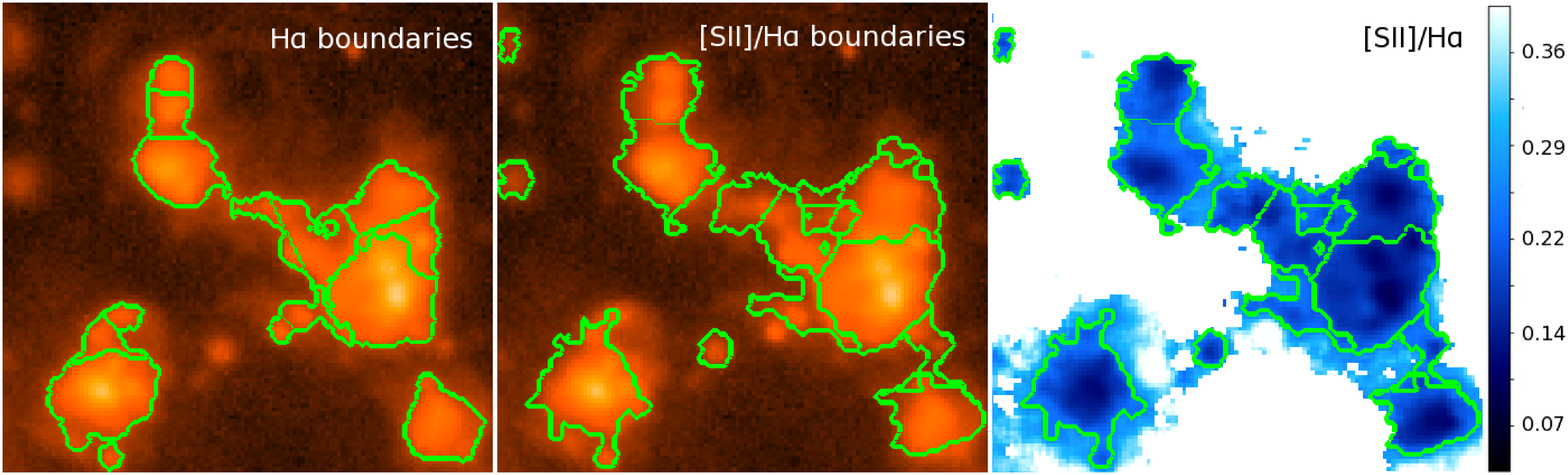}
\caption{Top: V-band image of NGC 628 with the H$\alpha$ maps for the three MUSE fields overlaid, with contours for the spiral arms (red).   H$\alpha$ identified HII regions using Clumpfind (green) cleanly select isolated HII regions and break up more crowded HII complexes into individual regions.  
Bottom: A sample comparison (white box, above) of the HII regions identified using H$\alpha$ line flux (left) or the [SII]/H$\alpha$ line ratio (center, right) to define region boundaries.  
[SII]/H$\alpha$ identified HII regions generally cover the same extent and more easily identify fainter regions, but are not as well able to divide clustered regions. 
\label{fig:NGC628}}
\end{figure*}

NGC 628 was observed using the Multi-Unit Spectroscopic Explorer (MUSE; \citealt{Bacon2010}) at the Very Large Telescope (VLT) in three positions (programme ID 094.C-0623 and ID 095.C-0473) centered on interarm regions (Figure~\ref{fig:NGC628}).  MUSE provides a 1\arcmin$\times$1\arcmin~ field of view with 0\farcs2 pixel size.  Observations were taken using the extended wavelength setting, covering 4650-9300\AA, with 0\farcs8 seeing.  Two northern and one southern pointings were observed with 845s and 990s exposures, respectively, in three 90\arcdeg~ rotations, alternated with separate sky exposures.   All reductions were carried out using the standard ESO pipeline version 1.2.1.  Emission lines H$\beta$ $\lambda$4861, [OIII] $\lambda$5007, H$\alpha$ $\lambda$6563, [NII] $\lambda$6548, [NII] $\lambda$6584, [SII] $\lambda$6717 and [SII] $\lambda$6531 are fit using LZIFU (Ho et al, in prep), tying all line kinematics and using MIUSCAT templates \citep{Vazdekis2012} for the underlying stellar continuum.   The typical spectral resolution for all lines fit ($<$7000\AA) is $\sim$2.75\AA (150 \kms).  We reach a 3$\sigma$ surface brightness sensitivity for H$\alpha$ of $1.5\times10^{-17}$~erg~s$^{-1}$~cm$^{-2}$~arcsec$^{-2}$. 

 In the H$\alpha$ map (not corrected for extinction) we geometrically identify HII region candidates using the 2D-Clumpfind algorithm \citep{Williams1994}, which finds local peaks in emission and extends them to lower flux levels.     
Peaks are identified using a lowest level at 25$\sigma$ (1.3 $\times$ 10$^{-16}$~erg~s$^{-1}$~cm$^{-2}$~arcsec$^{-2}$) with contour steps of 2$\sigma$.  This was found to provide thresholds sufficient to divide clustered areas and select isolated regions.  We omit as foreground star candidates the HII regions with more than 4 pixels where the stellar velocity is 300 km/s less than systemic for NGC 628.  
Using the H$\alpha$ surface brightness map we identify 391 HII regions (Figure~\ref{fig:NGC628}).  We correct  for dust attenuation using the H$\alpha$/H$\beta$ Balmer line decrement, assuming a Milky Way dust extinction law \citep{Cardelli1989} and R$_V$=3.1, and measure luminosities  10$^{36.1}$ erg~s$^{-1}<L_{H\alpha}<10^{39.0}$ erg~s$^{-1}$ (Figure~\ref{fig:hiireg}), with most below the break typically observed at L$_{H\alpha}=10^{38.7}$~erg~s$^{-1}$ \citep{Kennicutt1989HII}. 
All results are robust to variations in the exact H$\alpha$ boundaries identified. 

Decreased temperatures in the HII regions compared to the DIG results in a decreased [SII]/H$\alpha$ line ratio \citep{Madsen2006}, 
and provides an alternate diagnostic for identifying HII regions. Using Clumpfind on a [SII]/H$\alpha$ map we identify 319 HII regions  (Figure~\ref{fig:NGC628}, bottom), excluding foreground star candidates.  Here we have applied 1\arcsec~Gaussian smoothing, equivalent to the seeing, as the lower signal to noise ratio in [SII] results in irregular boundaries.  Equivalent smoothing of the H$\alpha$ line map does not significantly change the H$\alpha$ identified boundaries or HII region sizes.   As the HII regions are characterized by low [SII]/ H$\alpha$ values, Clumpfind was applied to an inverted map (multiplied by -1).  Peaks are identified using levels from 0.10, consistent with the expected pure HII region line ratio \citep{Madsen2006}, with contour steps of 0.025 in the flux ratio to a limiting level at 0.25.  As the flux ratio is decoupled from the line intensities, we found this step size to be the minimum we could use that still resulted in reasonable division of clustered regions.  However, in some regions this is insufficient to divide clearly distinct peaks (see Figure~\ref{fig:NGC628}).  Our level corresponding to the outer extent of the identified HII regions was  chosen to balance the observed thresholds with the decreasing signal in the [SII] line.  These boundaries are thus somewhat irregular, reflecting mainly the decreased line strength.  
   
As we wish to compare star formation across environments, it is important to use a spiral arm tracer directly linked to the underlying gravitational potential and not one biased by the star formation itself.  We use the mask from \cite{Foyle2010}, who use a smoothed 3.6$\mu$m map to locate the old stellar population, and Fourier decompose the resulting image.  However, their mask is limited to the central 2\arcmin~radius region and excludes the central 45\arcsec~of the galaxy.  As the  bulk molecular gas emission is strongly influenced by the gravitational potential \citep{Colombo2014},  we extend the spiral arms using HERACLES CO(2-1) maps \citep{Leroy2009} at 14\arcsec (500 pc) resolution (Figure~\ref{fig:NGC628}). Using the H$\alpha$ boundaries, we find 271 HII regions on the spiral arm and 120 in the interarm.  Using the [SII]/H$\alpha$ boundaries, we find 192 HII regions on spiral arms and 127 in the interarm.  Fewer regions are detected in the arm with this method as it is less effective at dividing highly clustered HII complexes (see Section~\ref{sec:siiha}).

\clearpage
%>>>>>>>>>>>>>>>>>>>>>>>>>>>>>>>>>Section<<<<<<<<<<<<<<<<<<<<<<<<<<<<<<<<<<<<<<<<<<<<
\section{Results}
Our IFU observations provide spectral coverage for a large number of HII regions over their full spatial extent. This allows us to go beyond the parameters obtained in imaging studies (luminosities, sizes) to a full characterization of the physical conditions. In particular, we can more accurately account and correct for diffuse contribution to H$\alpha$ emission, and test the assumptions traditionally used in these corrections. 

\subsection{HII region properties}
\label{sec:hiiprop}
We present in Figures \ref{fig:hiireg} and \ref{fig:siiha} a comparison of HII region properties for arm and interarm regions, using both HII region identification methods.  As the results are similar for both methods, we focus here on the arm/interarm comparison using the H$\alpha$ boundaries and discuss differences between the two methods in Section~\ref{sec:siiha}.

The range of HII region luminosities measured appears similar in both environments (a Kolmogorow-Smirnow test cannot distinguish between these two populations), but suffers from many biases.  As our fields are centered on interarm regions, we miss the brightest spiral arm HII regions.  In addition, our 35 pc scale resolution limits us to resolving the brightest regions $>$10$^{38}$~erg~s$^{-1}$   \citep{Gutierrez2011}.  Due to these biases we do not attempt to fit the luminosity function, but show our luminosity distribution in relation to the slope previously reported for NGC 628 \citep{Kennicutt1980}.  More extended spatial coverage is necessary to disentangle any changes between our novel IFU based luminosity function and previous narrow band imaging results.  

We convert the size of each region, in pixels, to an equivalent radius and find 20~pc~$<r_{eq}<100$~pc.  HII region size is related to the age \citep{Whitmore2011}, though we are only resolving the largest, and thus the oldest (6-8 Myr), of the HII regions and in this regime the size has little sensitivity as an age diagnostic.  The same range of sizes is seen in both environments, with nearly half of regions unresolved (r$\sim$35 pc), though this is sensitive to both the region identification method and the increased diffuse background on the spiral arms (see Section~\ref{sec:dig}).   

HII region metallicities have a well established radial dependence \citep{Zaritsky1994, Ho2015}, but the 2D variations are less well explored.  M101 has shown evidence for azimuthal gas-phase metallicity variations \citep{Kennicutt1996, Li2013}, however \cite{Cedres2002} found no difference between arm and interarm metallicities.  Conversely, in NGC 628 and NGC 6946, \cite{Cedres2012} found evidence of higher metallicity HII regions along the spiral arms.  
We calculate the gas phase oxygen abundance using the \cite{Dopita2016}  strong line diagnostic, which accounts for the ionization state without the necessity of the [OII] 3727 doublet.  
Results are similar using the \cite{PP04}  N2 diagnostic.  
Linear fits to both populations agree within the fit uncertainties, and are consistent with the gradient measured using temperature sensitive methods \citep{Berg2015}.  %This is expected as the enrichment of the gas should be similar at similar radii. 
For the spiral arm HII regions we fit the radial metallicity gradient as
\begin{multline}
12+Log(O/H)\\  	 
= (8.790 \pm 0.017) + (-0.558 \pm 0.064) \times R(dex~R_{25}^{-1}).
\end{multline}
For the interarm HII regions we fit the radial metallicity gradient as 
\begin{multline}
12+Log(O/H)\\  	 
= (8.780 \pm 0.022) + (-0.422 \pm 0.101) \times R(dex~R_{25}^{-1}).
\end{multline}

[OIII]/H$\beta$, which correlates with the ionization parameter, is also known to vary radially \citep{Rosales-Ortega2011}.  
Little work has been done investigating environmental dependences on this parameter, however \cite{Cedres2002} observed no difference in this ratio between arm and interarm HII regions within M101.  
We observe no difference between the arm and interarm populations, and linear fits to both populations agree within the fit uncertainties.

\subsection{Diffuse ionized gas}
Not all H$\alpha$ emission observed arises from HII regions.  A large fraction of H$\alpha$ emission in galaxies can arise from the diffuse ionized gas (see reviews by \citealt{Mathis2000} and \citealt{Haffner2009}).  Quite different values for the diffuse fraction have been found for interarm regions, from 0\% for NGC 247 \citep{Ferguson1996} to 100\% for the central regions of M51 \citep{Blanc2009}.  Given the sparse distribution and lower luminosities of HII regions in the interarm regions and the low surface brightness of the DIG, the diffuse fraction may depend significantly on the resolution and surface brightness sensitivity of the observations.  

We quantify the total contribution from diffuse emission to the H$\alpha$ flux in the arm and interarm by summing all flux in each environment, and comparing it to the total flux in each environment that is outside of the identified HII regions.  
We measure a 17~$\pm$~2\% diffuse fraction in the  arm, which rises to 49~$\pm$~10\% in the interarm.  Close examination of the H$\alpha$ images reveals there is still some contribution from low-level interarm star formation to what we have identified as diffuse emission.  
Lowering the H$\alpha$ surface brightness threshold used in the Clumpfind algorithm to 15$\sigma$ (7.5 $\times$ 10$^{-17}$~erg~s$^{-1}$~cm$^{-2}$~arcsec$^{-2}$), which more completely identifies discrete structures, results in a decrease of the interarm diffuse fraction to 36~$\pm$~9\%, but also associates many extended DIG features in the arm with the large HII region complexes.

The [SII]/H$\alpha$ line ratio correlates with diffuse gas fraction, showing a narrow range of low values for emission purely from HII regions and increasing values depending on the relative flux contribution from the diffuse gas  \citep{Madsen2006, Blanc2009}.  We explore the [SII] $\lambda$6716+$\lambda$6731/H$\alpha$ line ratio for our HII regions and find a difference between the arm and interarm populations (Figure~\ref{fig:siiha}). The interarm regions span a narrower range of values, and are generally more consistent with pure HII regions.  The arm regions show an anti-correlation with the H$\alpha$ luminosity, indicating a larger diffuse fraction in fainter HII regions.  This was seen clearly in M51 by \cite{Blanc2009}, however, the difference between our arm and interarm HII regions suggests differing levels of contamination for the two environments.

%>>>>>>>>>>>>>>>>>>>>>>>>>>>>>>>>>Section<<<<<<<<<<<<<<<<<<<<<<<<<<<<<<<<<<<<<<<<<<<<
\section{Discussion}
The combination of spectral diagnostics mapped at high spatial resolution enables us to investigate some of the assumptions made when isolating the H$\alpha$ emission that is associated with star formation.  We model the impact of DIG contamination within our HII regions, and explore the novel use of [SII]/H$\alpha$ for identifying HII region boundaries.  Finally, we measure the impact the DIG correction could have on star formation studies.

%>>>>>>>>>>>>>>>>>>>>>>>>>>>>>>>>>Subsection<<<<<<<<<<<<<<<<<<<<<<<<<<<<<<<<<<<<<<<<<<<<
\subsection{Diffuse background within HII regions and the distribution of DIG}
\label{sec:dig}
Given the variation in [SII]/H$\alpha$ seen between HII regions, we investigate the use of this ratio to measure the fraction of H$\alpha$ emission due to diffuse ionized gas contributing as a background to the HII regions.   Following \cite{Blanc2009}, we model the H$\alpha$ flux as arising from two sources, 

\begin{equation}
\label{eqn:chii}
\begin{array}{ccl}
f(H\alpha) & = & f(H\alpha)_{H II}+f(H\alpha)_{DIG}\\\;\\
           	& = & C_{H II}f(H\alpha)+C_{DIG}f(H\alpha)
\end{array}
\end{equation}
where $C_{DIG}$ is the fraction of total flux arising from the DIG and $C_{DIG}=1-C_{HII}$.  The observed [SII]/H$\alpha$ ratio is then
\begin{equation}
\label{eqn:chii2}
\frac{[SII]}{H\alpha} = C_{HII}\left(\frac{[SII]}{H\alpha}\right)_{HII}+C_{DIG}\left(\frac{[SII]}{H\alpha}\right)_{DIG}.
\end{equation}

We measure the typical [SII]/H$\alpha$ ratio in the arm and interarm, taking the median value for the HII regions and using an integrated spectrum for diffuse regions (Table~\ref{tab:siiha}).  
([SII]/H$\alpha$)$_{DIG}$ is similar between the arm and interarm, supporting our claim that we identify the bulk of the HII regions within both environments, and is systematically higher than ([SII]/H$\alpha$)$_{HII}$.  

\begin{figure*}
\centering
\includegraphics[width=5in,trim=0cm 1cm 0cm 0cm,clip]{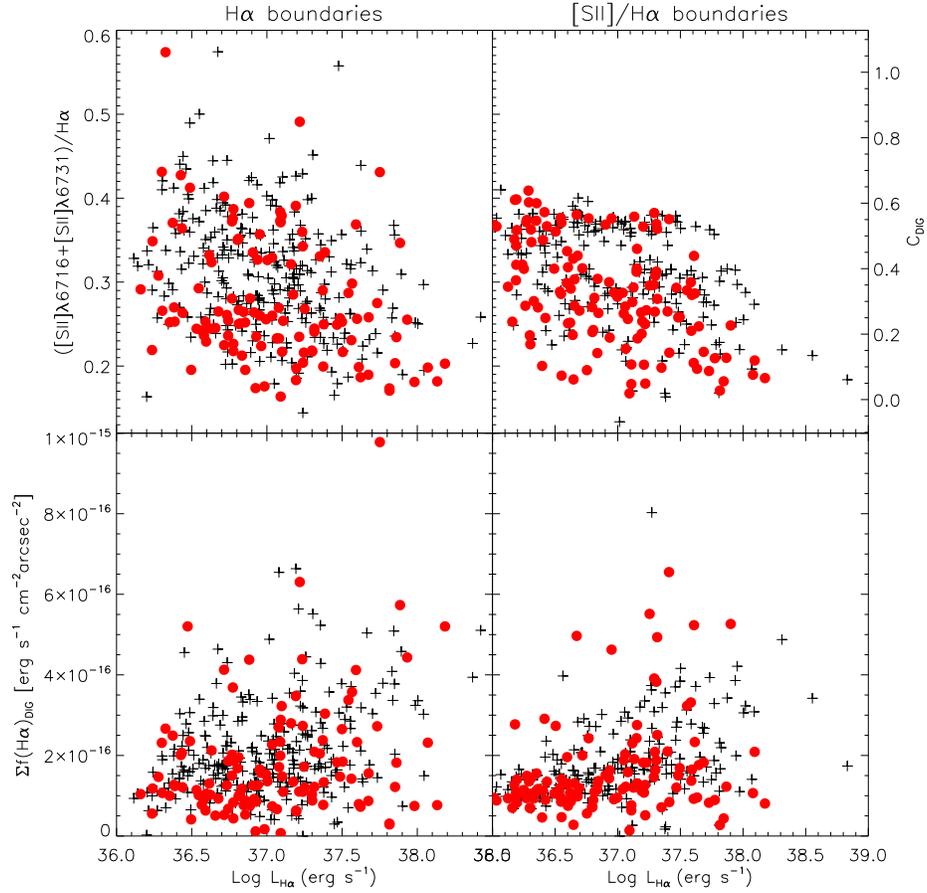}
\caption{Using [SII]/H$\alpha$ to constrain DIG emission for the arm (black) and interarm (red) HII regions, comparing HII regions identified using H$\alpha$ emission (left) or [SII]/H$\alpha$ line ratios (right). Top: [SII]/H$\alpha$ ratio as a function of HII region luminosity.  Typical uncertainty is 0.04.  Corresponding diffuse gas fraction (C$_{DIG}$) is shown on the right axis.  Interarm HII regions show ratios more consistent with pure star formation, while arm regions suffer from a larger diffuse gas contamination. Bottom: H$\alpha$ surface brightness contamination due to the diffuse gas background to each HII region as a function of HII region luminosity.    
\label{fig:siiha}}
\end{figure*}

The [SII]/H$\alpha$ ratios within the HII regions show a range of values (Figure~\ref{fig:siiha}), with the interarm HII regions having consistently lower ratios.  There are a number of possible explanations for this apart from a difference in DIG contribution.  [SII]/H$\alpha$ decreases with increasing ionization parameter, lower metallicity, decreasing spectral hardness (age) and increased ionizing radiation escape fraction.  We have constrained the first two conditions and found no difference with environment (Section \ref{sec:hiiprop}), and have no way to constrain the last two conditions.   These emission lines are near enough in wavelength that the effect of dust reddening is negligible.

Assuming the variations in [SII]/H$\alpha$ are due to DIG contamination, we quantify this within each HII region.  Following Equation \ref{eqn:chii2} we calculate $C_{DIG}$, assuming ([SII]/H$\alpha$)$_{HII}$ = 0.16, the minimum value for our sample, and ([SII]/H$\alpha$)$_{DIG}$ = 0.51, the average from Table~\ref{tab:siiha}.  Use instead of the single [SII]$\lambda$6716 \AA~line, as in \cite{Madsen2006}, gives similar results and a ratio ([SII]$\lambda$6716/H$\alpha$)$_{HII}$ = 0.10 that is consistent with Milky Way HII regions \citep{Madsen2006}.  We find a median diffuse fraction that is nearly twice as high in the HII regions on the arm compared to the interarm  (44$\pm$11\% vs 28$\pm$10\%), and an anti-correlation with L$_{H\alpha}$ (Figure~\ref{fig:siiha}).    

We then calculate f(H$\alpha$)$_{DIG}$, and based on the spatial extent convert this to a DIG surface brightness.  For the DIG regions, we use an integrated spectrum to directly measure the extinction corrected H$\alpha$ surface brightness.  All values are listed in Table~\ref{tab:siiha}.  We observe a higher DIG surface brightness in the arms compared to the interarm, and a higher DIG surface brightness contribution within the HII regions compared to the surrounding area.  This is expected if continuum radiation leaked from HII regions contributes significantly to photoionizing the DIG \citep{Zurita2000}, and suggests we may be systematically underestimating the ionizing radiation emitted by the HII region as we have not accounted for these leaked photos.

This difference in DIG background level could bias the measured sizes on spiral arms to be larger than similar sized HII regions in the interarm when using a H$\alpha$ surface brightness threshold.  This could also bias the luminosity limit  for detecting HII regions within the arm and interarm.  

%>>>>>>>>>>>>>>>>>>>>>>>>>>>>>>>>>Subsection<<<<<<<<<<<<<<<<<<<<<<<<<<<<<<<<<<<<<<<<<<<<
\subsection{HII region identification by physical conditions.}
\label{sec:siiha}

The multiple line maps obtained from our optical IFU spectroscopy provide a novel opportunity to revisit established methods for identifying HII region boundaries.  The [SII]/H$\alpha$ line ratio provides a method that is based on variations in temperature rather than line intensity.   
 We find that the HII regions generally cover the same extent, and we more easily identify fainter regions, but are less successful at dividing clustered regions (Figure~\ref{fig:NGC628}).   The region boundaries are more irregular due to the lower intensity of the [SII] line.  

Our catalog of [SII]/H$\alpha$ identified HII regions results in a realistic range of luminosities and sizes (Figure~\ref{fig:hiireg}). 
The arm/interarm populations appear more similar in their size distribution compared to HII regions defined by the H$\alpha$ intensity alone.  
The diffuse fractions in the arm and interarm for these [SII]/H$\alpha$ boundaries are 21~$\pm$~2\% and 55~$\pm$~11\%, respectively.  This is consistent with what was found with the H$\alpha$ boundaries, supporting our claim that these temperature-identified HII regions produce similar overall recovery of HII region emission as the traditional intensity-identified method.

Within the HII regions, the [SII]/H$\alpha$ line ratios are restricted, by definition, to lower values (Figure~\ref{fig:siiha}).  
We still see on average lower DIG fractions for interarm HII regions, suggesting this is not just an H$\alpha$ boundary issue biased by increased DIG background on the arms.  The median diffuse background surface brightness contribution within HII regions is similar ($1.62\times10^{-16}$ and $1.19\times10^{-16}$~erg~s$^{-1}$~cm$^{-2}$~arcsec$^{-2}$ in the arm and interarm, respectively), and follows the same trend with environment. 

As maps of this line ratio will be available for most optical IFU studies, it provides a new method to distinguish HII region boundaries, and will be essential to constructing H$\alpha$ line emission maps cleaned of all contaminants for use in star formation studies. 

\clearpage
%>>>>>>>>>>>>>>>>>>>>>>>>>>>>>>>>>Subsection<<<<<<<<<<<<<<<<<<<<<<<<<<<<<<<<<<<<<<<<<<<<
\subsection{Kennicutt-Schmidt law at 35pc and the role of spatial averaging}

We explore variations in the Kennicutt-Schmidt law \citep{Schmidt1959, Kennicutt1989, Kennicutt1998} to understand how the efficiency of star formation changes with galactic environment.    \cite{Brinchmann2013} showed that optical spectroscopy alone can be used to estimate the total gas mass surface density within a factor of two on kpc scales, and we use here a simplified model based on recent empirical results at smaller spatial scales.   

We use dust attenuation as measured from the H$\alpha$/H$\beta$ Balmer line decrement, assuming a Milky Way dust extinction law \citep{Cardelli1989} and R$_V$=3.1, to estimate the dust mass surface density (see \citealt{Kreckel2013}).  
Assuming a fixed dust to gas ratio of 0.01  \citep{Sandstrom2013} we convert this to a total gas surface density.   
Note that this method suffers a surface brightness bias, as we can only measure high extinction (and therefore high gas mass) on high surface brightness (and therefore high SFR) HII regions.  
We use the extinction corrected H$\alpha$ luminosity and HII region sizes to calculate the SFR surface density \citep{Kennicutt1998}.  Note that due to its stochastic nature, SFR is probably not well defined for these $\sim$35~pc spatial scales, when selecting individual HII regions.  We are also potentially systematically underestimating our SFR as we have not accounted for leaked photos that contribute to ionizing the DIG.

\begin{figure*}
\centering
\includegraphics[width=3.3in]{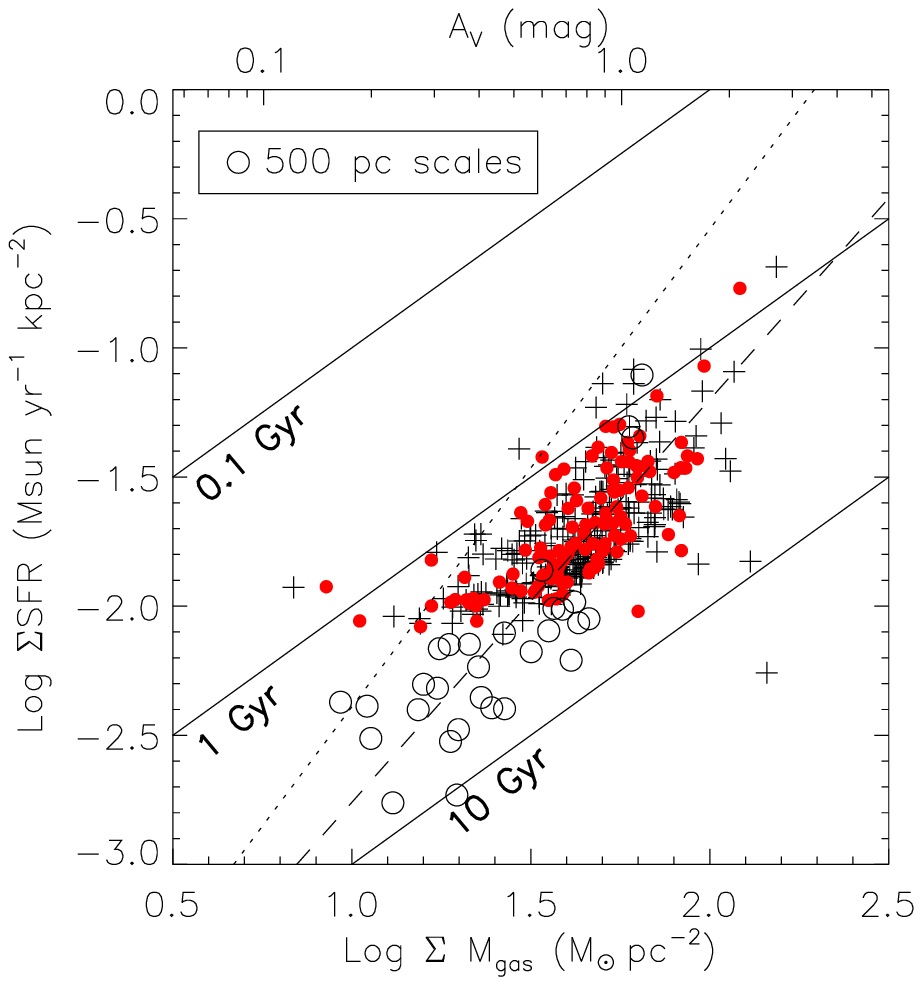}
\includegraphics[width=3.3in]{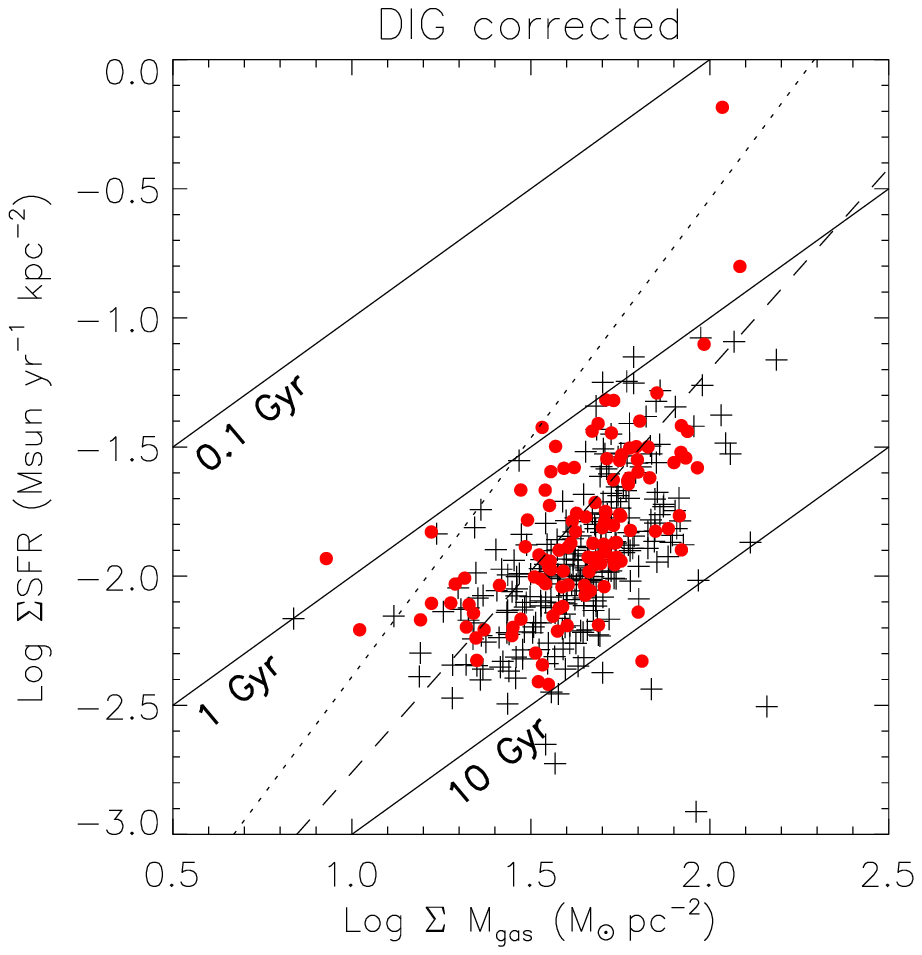}
\caption{The total gas Kennicutt-Schmidt law at 35 pc scales.  Constant depletion times of 10$^8$, 10$^9$ and 10$^{10}$ yr are shown as solid lines.  Left: Arm (black) and interarm (red) HII regions are consistent with \cite{Kennicutt2007} (dashed) and \cite{Bigiel2008} (dotted), though those studies were limited to larger (350-500~pc) spatial scales.  Binning our regions to 500~pc scales (open circles) results in longer depletion times.   Right: Including a DIG correction shifts all regions to longer depletion times by 40\% for the interarm and 62\% for the arm, and steepens the relation.
\label{fig:ks}}
\end{figure*}

Compared to previous sub-kpc scale SF studies using total gas \citep{Kennicutt2007, Bigiel2008}, we see good agreement with the established relations considering the significant assumptions that go into our model (Figure~\ref{fig:ks}).   
Calculating a gas depletion time, we find similar values for arm and interarm of $2\times10^9$ yrs.  
We see no difference between arm and interarm regions.  \cite{Foyle2010} also found no evidence for differing gas depletion times at 250-600pc scales.  
Other nearby galaxies show very different results.  NGC 6946, a more flocculent spiral galaxy, shows increased star formation efficiency (lower depletion times) on spiral arms \citep{Rebolledo2012}, and M51 exhibits a large scatter in the star formation efficiency along its spiral arms \citep{Meidt2013}.

The gas depletion times we estimate here are significantly longer than the lifetimes typically measured for GMCs \citep{Meidt2015}.  However, if clouds are destroyed via shear or feedback on shorter (20-30 Myrs) timescales they must also be created in order to maintain the present rate of star formation and the present efficiency.   This is consistent with a constant cycling of molecular material between bound, cloud-like objects and a more diffuse state \citep{Meidt2015}. 

We observe good agreement with the results of  \cite{Schruba2010} when they explored the role spatial resolution plays in calculated depletion times.  As we are targeting SF regions, increasing our aperture to cover larger spatial scales will necessarily introduce regions without ongoing star formation, and lower the SFR surface density.   Spatial averaging over 500pc scales results in longer depletion times (Figure~\ref{fig:ks}, left). However, given that we should be biasing our results to high SFRs as we are purposefully selecting only HII regions, our gas depletion times are an order of magnitude longer than those found by \cite{Schruba2010} at 75 pc scales with similar biases in M33.  Direct observation of the molecular gas are necessary to confirm the environmental trends that are suggested here in NGC 628.  

Including a correction for the diffuse gas contamination of each HII region results in decreased H$\alpha$ luminosities and hence lower SFRs and longer depletion times of $\sim$1~Gyr (Figure~\ref{fig:ks}, right).  This is on average a 62\% (40\%) increase in the spiral arms (interarms), and anti-correlates with HII region luminosity, thus steepening the Kennicutt-Schmidt law.  Given the relatively large diffuse contribution, particularly within spiral arms, this correction factor must be taken into account when calculating H$\alpha$ SFRs.
For a global spectrum, this DIG correction results in a 56\% change, a significant correction to consider for unresolved (high redshift) sources.

%>>>>>>>>>>>>>>>>>>>>>>>>>>>>>>>>>Subsection<<<<<<<<<<<<<<<<<<<<<<<<<<<<<<<<<<<<<<<<<<<<
\section{Implications}
Recent results have shown that there is no `universal' giant molecular cloud (GMC), with cloud properties varying significantly between arm, interarm, upstream and downstream regions \citep{Colombo2014}.  However, our study suggests that while the birthplace properties may change, the resulting HII regions appear largely indistinguishable.  This supports recent statistical studies comparing spiral arm and interarm HII region luminosity functions, which largely show no difference \citep{Scoville2001, Azimlu2011, Gutierrez2011}.
 
\begin{figure*}
\centering
\includegraphics[width=4.5in,trim=1.5cm 1cm 0.9cm 0.5cm,clip]{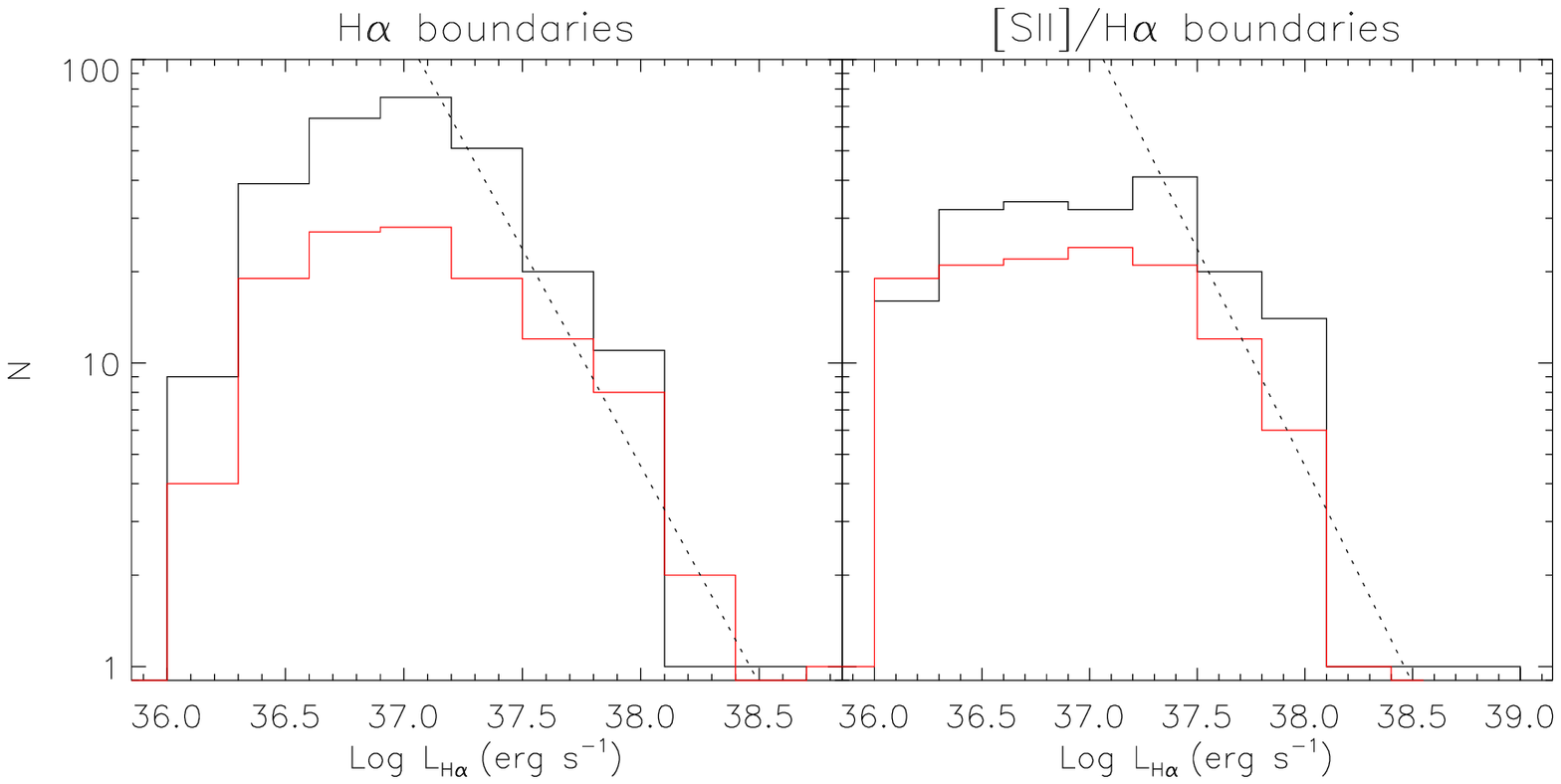}
\includegraphics[width=4.5in,trim=1.5cm 1cm 0.9cm 0.5cm,clip]{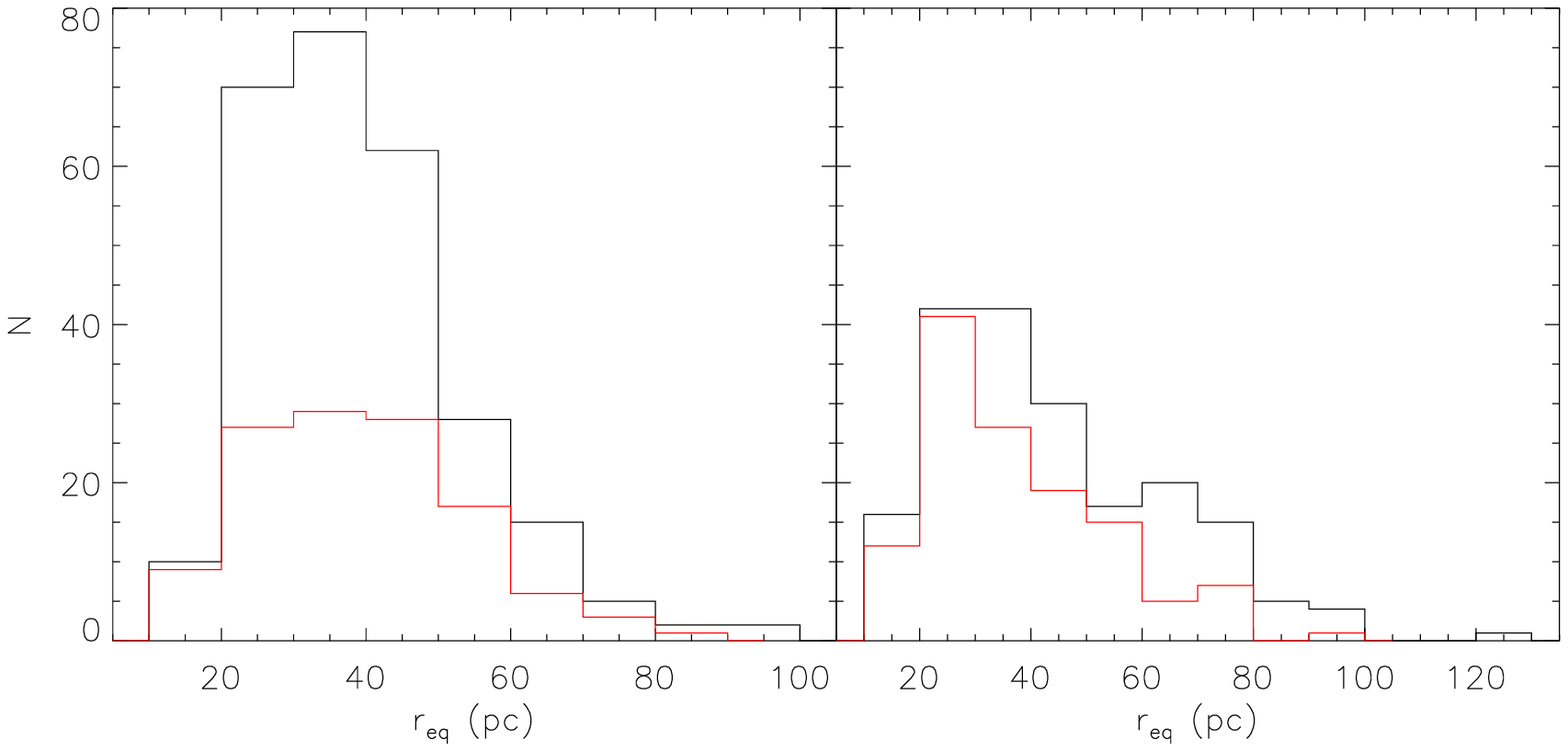}
\includegraphics[width=4.5in,trim=1.5cm 1cm 0.9cm 0.5cm,clip]{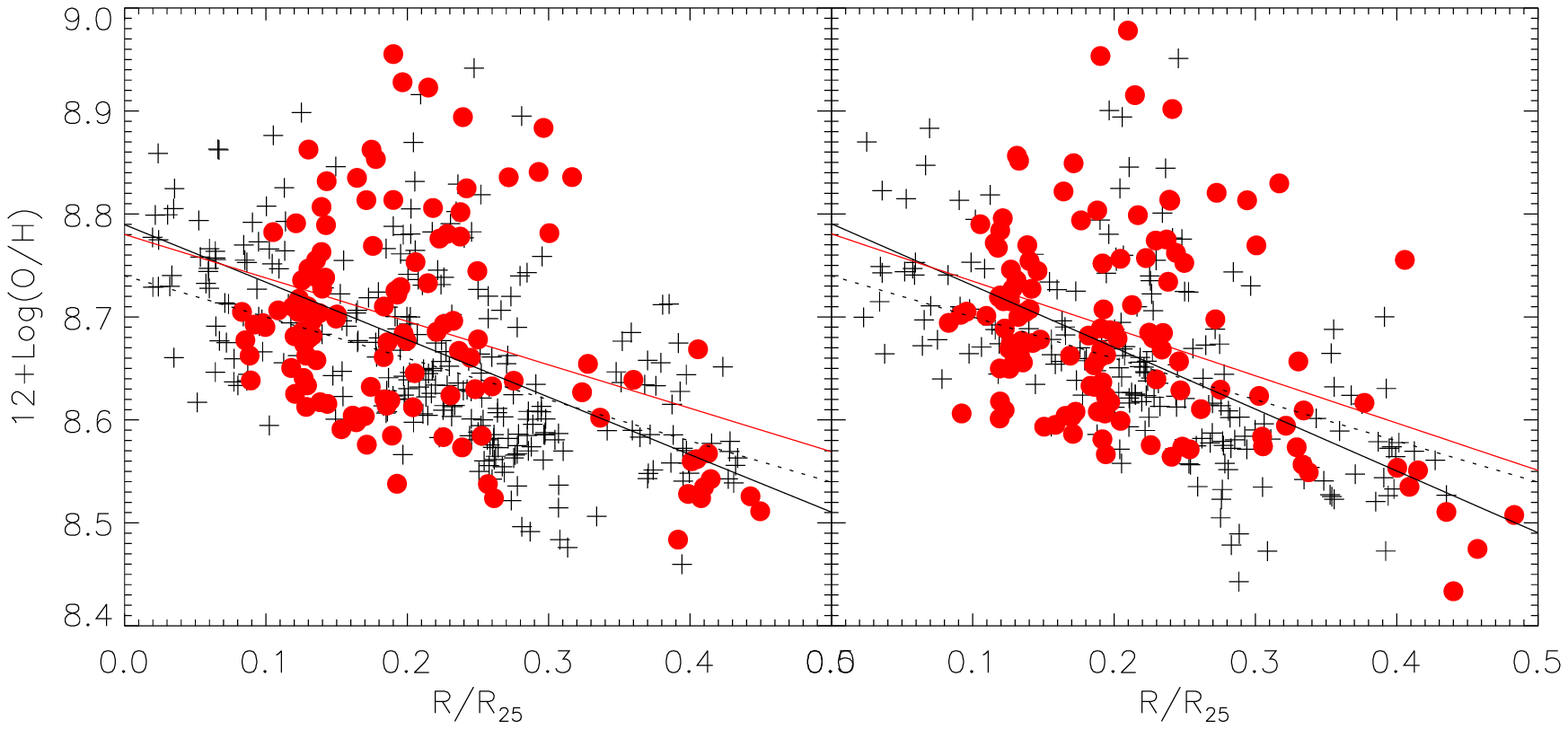}
\includegraphics[width=4.5in,trim=1.5cm 1cm 0.9cm 0.5cm,clip]{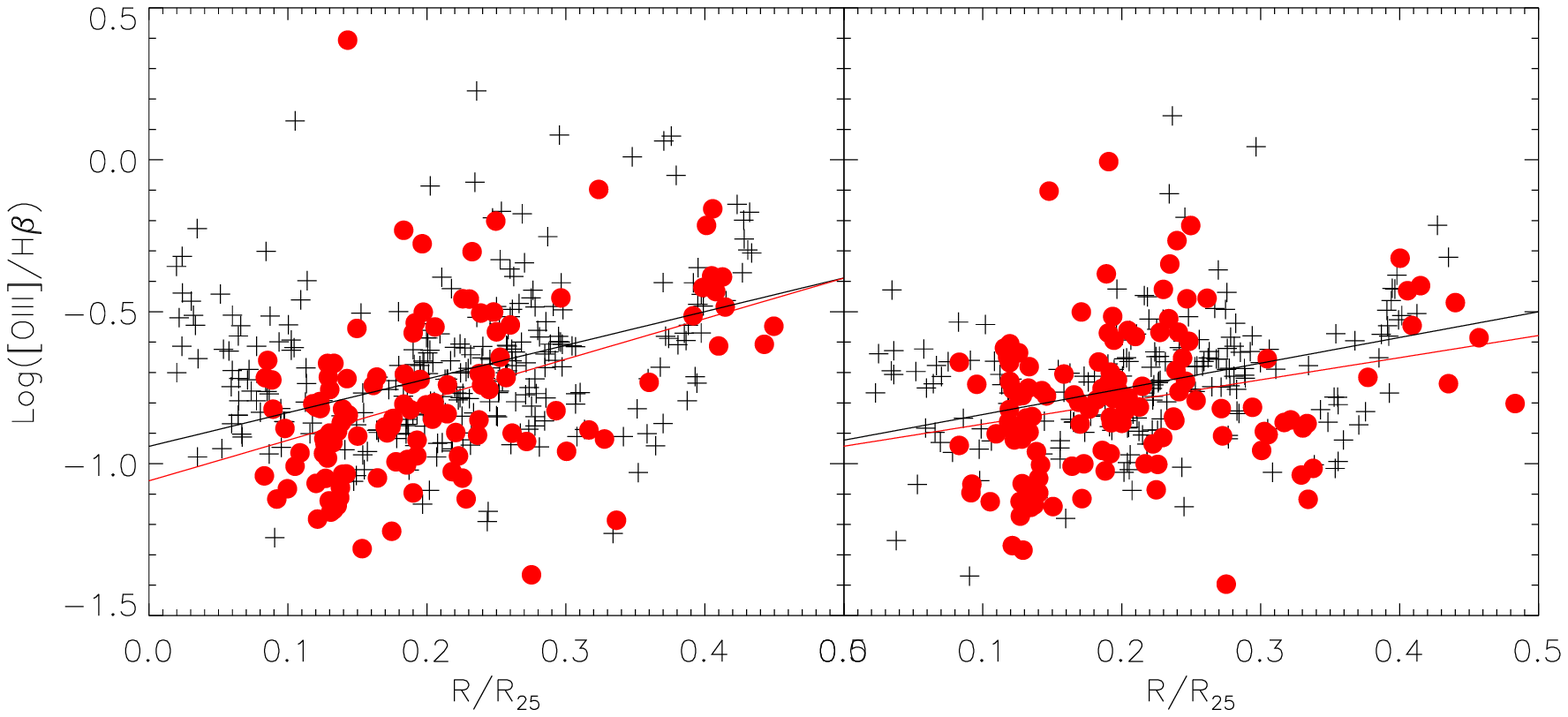}
\caption{A comparison of HII region properties from the arm (black) and interarm (red) regions, using HII regions identified via H$\alpha$ intensity (left) or [SII]/H$\alpha$ line ratios (right).  Top panel: The range of HII region luminosities measured is similar between environments, and has a shallower slope compared to previous work (dotted line; \citealt{Kennicutt1980}).  Upper middle panel: The distribution of r$_{eq}$ is also similar between environments, but depends strongly on the HII region identification method.  Lower middle panel: The metallicity follows the established radial gradient (dotted line; \citealt{Berg2015}),  typical uncertainty is 0.1 dex.  Bottom panel: [OIII]/H$\beta$, which correlates with ionization parameter, also varies radially.  Typical uncertainty is 0.1 dex.
No significant difference is seen in these parameters when comparing the arm and interarm regions.
\label{fig:hiireg}}
\end{figure*}

Spiral arm GMCs are typically more massive, and are therefore expected to form a larger number of massive stars, inject more mechanical energy and ionize the gas more strongly.  Given that we see no difference between the ionization state of arm and interarm HII regions (Figure \ref{fig:hiireg}),  either the efficiency with which they form is lower in the arms or the gas in the arms is less affected by feedback (more porous or fragmented).

Alternately, our result might reflect the fact that the more extended envelope of the GMC is sensitive to local environmental effects while the star formation process that occurs in the dense clumps and cores is largely decoupled from the cloud as a whole.  While CO cloud properties may regulate the number of dense clumps that form, the process by which the densest gas forms stars (and creates an HII region) is universal.  This is consistent with findings that HCN gas surface density is linear with $\Sigma_{SFR}$ and largely independent of local conditions, while CO gas surface density shows larger variation \citep{Gao2004, Usero2015, Bigiel2016}.

Future resolved, cloud-scale study of the molecular and dense gas in NGC 628 will directly test these scenarios.

\acknowledgments
We would like to thank the referee for their helpful comments.  We thank Kelley Foyle for providing  arm/interarm masks, and I-Ting Ho for early use of LZIFU.  KK acknowledges grant KR 4598/1-2 and SM acknowledges grant SCHI 536/7-1 from the DFG
Priority Program 1573.  GB is supported by CONICYT/FONDECYT, Programa de Iniciacion, Folio 11150220.  BG gratefully acknowledges the support of the Australian Research Council as the recipient of a Future Fellowship (FT140101202).   AH acknowledges support from the Centre National d'Etudes Spatiales (CNES). 
Based on observations made with ESO Telescopes at the La Silla Paranal Observatory under programme ID 094.C-0623 and ID 095.C-0473.

\clearpage

\begin{deluxetable*}{l c c}
\tablecaption{ [SII]/H$\alpha$ ratios for arm and interarm regions.  \label{tab:siiha}}
\tablehead{
\colhead{} & 
\colhead{[SII]/H$\alpha$} &
\colhead{f(H$\alpha$)$_{DIG}$ }
\\
\colhead{} & 
\colhead{} & 
\colhead{[erg~s$^{-1}$~cm$^{-2}$~arcsec$^{-2}$]} 
}
\startdata
HII regions \\
\hline
arm\tablenotemark{a} 				&	0.32	&	1.78 $\pm$ 0.10 $\times$ 10$^{-16}$\\
interarm\tablenotemark{a} 			&	0.26	&	1.35 $\pm$ 0.08 $\times$ 10$^{-16}$\\
\hline
DIG  \\
\hline
arm\tablenotemark{b} 				&	0.47		&	1.29 $\pm$ 0.12 $\times$ 10$^{-16}$\\	
interarm\tablenotemark{b} 			&	0.54 		&	0.86 $\pm$ 0.11 $\times$ 10$^{-16}$	
\enddata
\tablenotetext{a}{Median value found within HII regions, f(H$\alpha$)$_{DIG}$ calculated following Equation \ref{eqn:chii}.}
\tablenotetext{b}{Measured from integrated DIG spectrum.}
\end{deluxetable*}

\end{document}